\documentclass[runningheads]{llncs}

\usepackage[numbers]{natbib} 
\usepackage{graphicx}
\usepackage[linesnumbered,ruled,noend]{algorithm2e}
\usepackage{algpseudocode}
\usepackage{comment}
\usepackage{url}
\begin{document}

\title{Understanding and Predicting Characteristics of Test Collections in Information Retrieval}

\titlerunning{Understanding and Predicting Characteristics of Test Collections in IR}

\author{Md Mustafizur Rahman\inst{1} \and Mucahid Kutlu\inst{2} \and Matthew Lease\inst{1}}
\institute{University of Texas at Austin \and TOBB  University of Economics Technology\\
\email{\{nahid,ml\}@utexas.edu,  m.kutlu@etu.edu.tr}}
\maketitle

\begin{abstract}
Research community evaluations in information retrieval, such as NIST's Text REtrieval Conference (TREC), build reusable test collections by pooling document rankings submitted by many teams. Naturally, the quality of the resulting test collection thus greatly depends on the number of participating teams and the quality of their submitted runs. In this work, we investigate: i) how the number of participants, coupled with other factors, affects the quality of a test collection; and ii) whether the quality of a test collection can be inferred prior to collecting relevance judgments from human assessors. Experiments conducted on six TREC collections illustrate how the number of teams interacts with various other factors to influence the resulting quality of test collections. We also show that the reusability of a test collection can be predicted with high accuracy when the same document collection is used for successive years in an evaluation campaign, as is common in TREC.


\end{abstract}

\section{Introduction}

Evaluation of information retrieval (IR) systems in the Cranfield paradigm \cite{cleverdon1967cranfield} relies on the construction of reusable document {\em test collections} \cite{sanderson2010test}. A test collection consists of a set of documents to be searched, a set of search {\em topics} (expressed as user input queries that correspond to underlying user information needs), and judgments of document relevance to different topics. Typically a test collection is constructed by organizing a {\it shared-task} wherein the organizers (e.g., NIST TREC) provide a document collection and a set of topics developed by experts, then ask the participating teams to submit a ranking of the (predicted) most relevant documents (i.e., {\it runs}) for each topic. Subsequently, the documents to be judged are selected from the teams' document rankings, 
canonically via {\it pooling} \citep{sparck1975report} the set of top-ranked documents across all team submissions. 

%

In this work, we conduct experiments to shed light on the specific impact that the number of participating teams has on the quality of pooling-based test collections, particularly their {\it reusability}. We investigate the following two key research questions. {\bf RQ-1)} i) how does the number of participants interact with other factors to influence the quality of a test collection? {\bf RQ-2)} Can we predict the quality of a test collection prior to collecting relevance judgments?   

For RQ-1, our experiments vary the number of participating teams by down-sampling submissions from past TREC evaluations in order to construct simulated test collections. Then we analyze the interaction between the number of submissions and other factors (e.g., the number of topics, collection size, etc.) on resultant test collection quality.  For RQ-2, we develop a model to predict test collection quality using only the number of participants, the number of topics, collection size, and pool depth. We analyze the generalization performance of the prediction model using our designed ``Leave-one-test-collection-out'' setup. 

Experiments conducted over six TREC test collections yield the following findings. Firstly, when we have very few participating teams, increasing the pool depth improves reusability more than increasing the number of topics. Secondly, the size of the document collection and the types of runs play a crucial role when the number of participants is very small. Thirdly, the reusability of a test collection can be predicted with high accuracy when the test collections used for training and testing the model have the same underlying document collection. This means that the results in one year can be used to effectively forecast the quality of the test collections built for the track in following years. To ensure the reproducibility of our findings, we have made our source code\footnote{{https://github.com/mdmustafizurrahman/Understanding-and-Predicting-the-Characteristics-of-Test-Collections-in-Information-Retrieval}} publicly available. 
%







\section{Factors Impacting the Qualities of Test Collections}\label{sec_rel}

The quality of a test collection depends on a variety of factors, such as: i) the number of topics; ii) pool depth (assuming pool-based judging); iii)  the number of participants; iv) the collection size; and v) the types of runs (e.g., manual runs and automatic runs) and their quality. One might also consider vi) the target evaluation metric (e.g., MAP@1000, NDCG@10) in assessing how well a test collection supports reliable evaluation for a given retrieval task, as measured by a particular metric. Since constructing a test collection is expensive, {\em reusability} is desirable. Reusability is often measured by how a given run contributing to the pool would have been assessed if excluded from the pool. In this study, we focus on reusability as a key measure of the test collection quality.   

While considerable work \cite{kuriyama_pooling_nodate, KanoulasSIGIR2021, VoorheesTOIS2017} has investigated how the quality of a test collection is impacted by the above-mentioned factors, prior studies have not explored how the number of participants interacts with other factors. 

{\bf Number of topics.} \citet{sparck1976information} suggest that 250 topics can be acceptable, but 1000 topics are needed for reliable evaluation. \citet{voorhees_variations_2000} performs an empirical analysis on the TREC6 test collection and shows that system rankings computed based on 5 or 10 topics are relatively unstable, whereas a set of 25 or more topics produces a stable ranking of IR systems. \citet{buckley_evaluating_2000} calculate the error rate of various evaluation measures and find that for reliable evaluation the required number of topics should be at least 25. \citet{webber_statistical_2008} recommend that a set of 150 topics is required to statistically distinguish the performance of one IR system from other IR systems. \citet{zobel1998reliable} finds that a set of 25 topics can reasonably predict the performance of IR systems on a separate set of 25 topics. 

{\bf Pool depth}. Prior work has also studied the trade-off between collecting fewer judgments with more topics (i.e.,\ Wide and Shallow (WaS) judging) vs.\ more judgments with fewer topics (i.e.,\ Narrow and Deep (NaD) judging). \citet{carterette_evaluation_2008} report on TREC Million Query track and conclude that WaS judging produces a more reliable evaluation of IR systems than NaD judging. \citet{kutlu_intelligent_2018} find that NaD judging is preferred to WaS judging if we consider intelligent topic selection or other hidden costs of shallow judging, such as topic creation time and noisier judgments.  \citet{voorhees_building_2018}  investigates the impact of varying pool depth on the reusability of a test collection. 

{\bf Run types.} To see how different types of runs (e.g., manual and automatic) impact collection quality,  \citet{buttcher2007reliable} adapt the ``leave-one-group-out"\citep{zobel1998reliable} experiment by removing all unique documents contributed by the manual runs from the pool. The authors find that their setup ranks the runs differently than found in the original TREC 2006 Terabyte task.

{\bf Collection size.} \citet{hawking2003collection} study the effect of the document collection size on the Very Large Collection 
track \citep{hawking1998overview} and observe  a higher value of P@20 for runs in the larger collection. Interestingly, they find no change in value  between the large and small document collection when the runs are ranked based on MAP.

\section{The Impact of Varying the Number of Participants}\label{sec_exp_1}


Our experimental design for analyzing the impact of the number of groups is shown in  \textbf{Algorithm \ref{algorithm:ourmethod_I}}. 
First, we construct the original qrels ($Q_o$) using runs of all participant groups ($G$) 
\textbf{[Line 2]}. Then we evaluate all runs using this original qrels ($Q_o$) in terms of a ranking metric (e.g., MAP@1000 and NDCG@10) and store the ranking of runs in $E$ \textbf{[Line 3]}. 
Next, we change the group number, $g$,  from $1$ to $|G|$ with an increment of $1$ at each iteration to create test collection with varying number of participants \textbf{[Line 5]}. At each iteration, we randomly sample $g$ number of groups ($\hat{G_i}$) from the set of groups $G$ \textbf{[Line 7]} and construct the simulated test collection (i.e., qrels) $Q_g$ using only the participants in $\hat{G_i}$ {\bf[Line 8]}. Then, we evaluate all participating runs in set $G$ by using simulated test collection $Q_g$ in terms of a ranking metric (e.g., MAP@1000, NDCG@10, etc.) and store these new ranking of runs in $E_g$ {\bf[Line 9]}. We calculate the performance difference in terms of $\tau_{ap}$ \citep{yilmaz2008new} and Max Drop \citep{voorhees_building_2018} (i.e., the maximum drop in a run's rank,  between the original ranking of runs $E$ and the ranking of runs obtained via the respective simulated test collection $E_g$) {\bf[Line 11]}. Note that we calculate average scores across different group samples for a particular parameter setup. In addition, we can also utilize Algorithm \ref{algorithm:ourmethod_I} to experiments with a varying number of topics and varying pool depth because it takes the set of topics $T$ and pool depth $P$ as inputs.


\begin{algorithm}[h]
    \SetAlgoLined
    \SetKwInOut{Input}{Input}
    \SetKwInOut{Output}{Output}

    \Input{Set of groups $G$ ~$\bullet$~ Number of samples for groups $N$ ~$\bullet$~ Set of topics $T$ ~$\bullet$~ Pool depth $P$}
    \Output{~~E, A set of performance score indexed by group number}
    
    $R \gets$ Total number of runs from all groups in $G$ \\
    $ Q_o \gets Construct\_Qrels(G, T, P) $  \Comment{official qrels}\\
    $E \gets Evaluate\_runs(R, Q_o)$ \Comment{Evaluate all runs with $Q_o$}\\
    $\hat{E} \gets \emptyset$ \Comment{keeps scores of systems with reduced qrels}

    \For{group no $g\gets1$ \KwTo $|G|$}{
        \For{sample number $i \gets1$ \KwTo $N$}{
            $\hat{G_i} \gets$ randomly sample g groups from $G$\\
            $Q_g \gets$ ConstructQrels($\hat{G_i}, T, P)$ \\
            $E_g \gets Evaluate\_runs(R, Q_g)$ \Comment{Evaluate all runs using  qrels $Q_g$}\\
            $\hat{E} \gets \hat{E} \cup E_g$
        }
    }

    \KwRet{Evaluate\_Performance\_Difference(E,$\hat{E}$)}
\caption{Experimental Design}
\label{algorithm:ourmethod_I}
\end{algorithm}

\subsection{Datasets}
\label{section:datasets}

We conduct our experiments on six TREC tracks and datasets: the 2013-2014 Web Tracks on ClueWeb12\footnote{\url{lemurproject.org/clueweb12}}, the 2006 Terabyte track on Gov2\footnote{\url{ir.dcs.gla.ac.uk/test_collections/gov2-summary.htm}}, and the 2004 Robust Retrieval Task (Robust'04),  the 1999 TREC-8 {\em ad hoc} track, the 1998 TREC-7 {\em ad hoc} track on TIPSTER disks  $4-5$\footnote{\url{trec.nist.gov/data/docs_eng.html}} (excluding the {\it congressional record}). \textbf{Table \ref{table:group_runs_data}} provides statistics about test collections we use. Later tracks have fewer participants than earlier tracks both in terms of the number of groups and the submitted runs. Later tracks also tend to use larger document collections, without commensurate increase in pool depth, leading to an increasing prevalence of relevant documents in judged pools, from  $\sim$5\% to $\sim$40\%.

\begin{table*}[!htb] 
\begin{center}
\scriptsize
\caption{Statistics about the test collections used in this study.}
\begin{tabular}{|l|c|c|c|c|c|c|r|r|r|r|}
\hline
\bf Track & \bf \#Groups & \bf \#Manual & \bf \#Auto  & \ \bf Pool & \bf Collection & \bf \#Topics  & \bf \#Docs & \bf \#Judged  & \bf \%Rel \\
 &  & \bf Runs & \bf Runs & \ \bf Depth &  &  &  &  &  \\
\hline

WT'14  &  9  &  4  &  26  &  25 &  ClueWeb12 & 50 & 52,343,021 & 14,432 & 39.2\%   \\
WT'13  &  13  &  3  &  31  &  10 and 20 & ClueWeb12 & 50 & 52,343,021 & 14,474 & 28.7\% \\
TB'06  &  20  &  19  &  61  &  50  & Gov2 & 50 & 25,205,179 & 31,984  & 18.4\%  \\
Adhoc'99  &  40  &  9  &  62  &  100 & Disks45-CR & 50 & 528,155 & 86,830 & 5.4\%  \\
Adhoc'98  &  41  &  16  &  68  &  100  & Disks45-CR & 50 & 528,155 & 80,345 & 5.8\% \\
Robust'04  &  14  &  0  &  110  &  100  & Disks45-CR & 249 & 528,155 &  311,410 & 5.6\% \\

\hline
\end{tabular}
\label{table:group_runs_data}
\end{center}
\end{table*} 
\vspace{-2px}

\subsection{Results and Discussion} \label{sec_results}

\subsubsection{Impact of Number of Topics.} We first consider how the number of topics interacts with the number of groups in relation to test collection quality. To explore this, 
we study Robust'04 using various subsets of its 249 topics, randomly sampling $m$ topics, with $m \in \{50, 100, 150, 200, 249\}$. Robust'04 topics can be categorized into 5 sets: 301-350 (Adhoc'97), 350-400 (Adhoc'98), 401-450 (Adhoc'99), 601-650 (Robust'03-Hard), and 651-700 (Robust'04). 
To ensure coverage over the different sets in our sampling, we apply stratified sampling. {\bf Algorithm \ref{algorithm:ourmethod_I}} implements our experimental setup for a given topic subset (we vary the topic subset outside of the algorithm).
The pool depth is set to $100$, and we evaluate MAP@1000 and NDCG@10. 
     
\begin{figure*}[!htb]
\centerline{\includegraphics[scale=0.50]{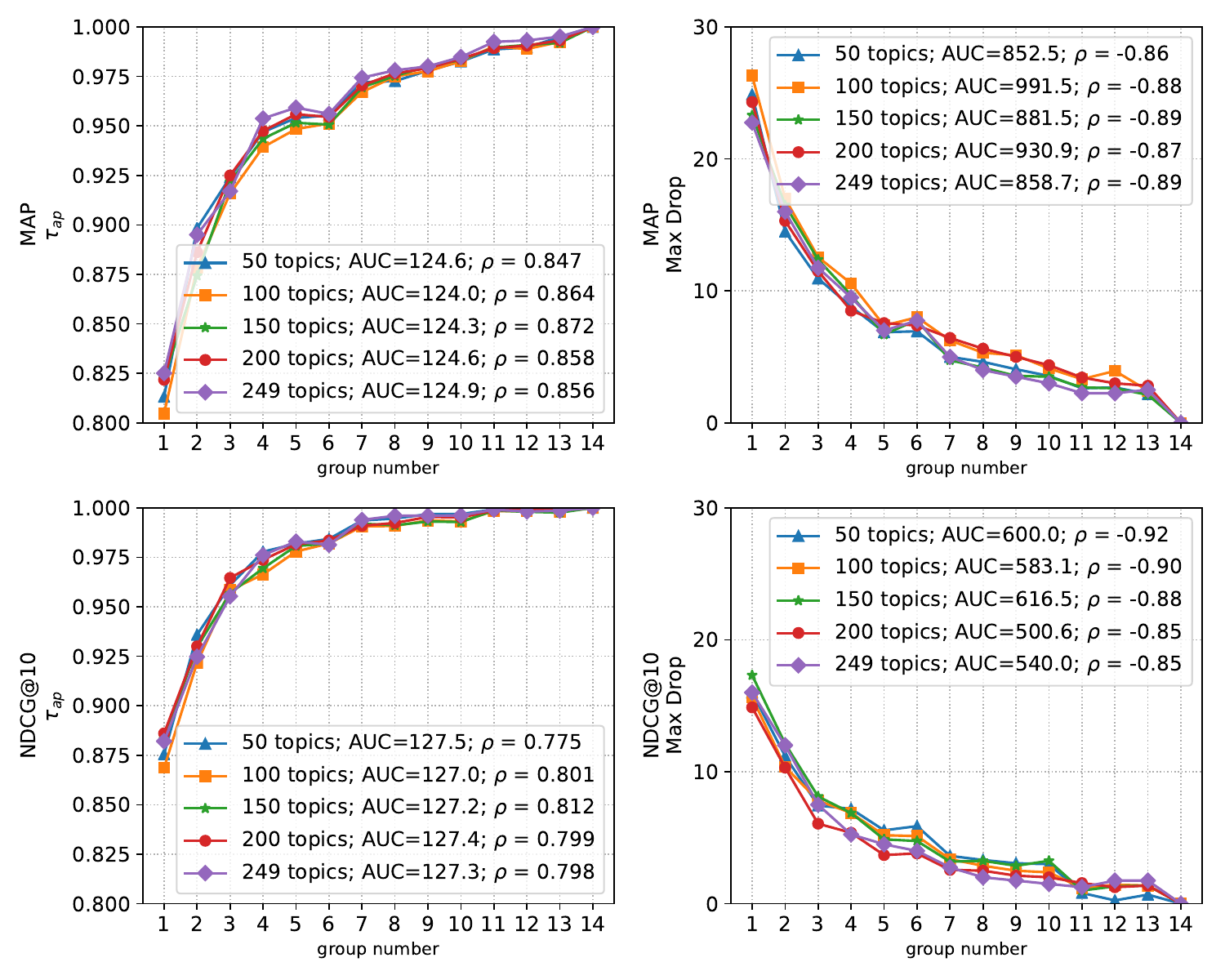}}\caption{$\tau_{ap}$ (first column), and Max Drop (second column) obtained by simulated test collections with a varying number of topics on the Robust'04 dataset. The x-axis represents the number of groups and the y-axis shows results when runs are ranked using MAP (top row) and NDCG@10 (bottom row).}
\label{Figure:topic_variation}
\end{figure*}

\textbf{Figure \ref{Figure:topic_variation}} shows $\tau_{ap}$ \citep{yilmaz_new_2008} (larger is better) and Max Drop \citep{voorhees_building_2018} (smaller is better) on the Robust'04 test collection using all runs. Each line shows an average value of the computed performance metric (e.g., $\tau_{ap}$) across $4$ different random samples (i.e., N is set to 4 in Algorithm  \ref{algorithm:ourmethod_I}) of groups when $m$ topics are randomly sampled from 249 topics.  
The first row presents $\tau_{ap}$ and Max Drop when the runs are ranked using MAP (i.e., MAP@1000) while the second row reports the same  metric when  runs are ranked using NDCG@10. We  report the Area under the Curve (AUC) for each of the line plots and the Pearson correlation ($\rho$) between the number of groups and the corresponding performance metrics.

Let us first consider when runs are ranked by MAP (Figure \ref{Figure:topic_variation}, first row). For $\tau_{ap}$ and any number of groups $g$, we do not see any significant difference in AUC when we down-sample to 50, 100, 150, or 200 topics vs.\ all 249 topics. Furthermore, we achieve a $\tau_{ap}$ correlation of $0.9$ using only $3$ participating groups, irrespective of the number of topics. We observe the same outcome when using NDCG@10 (Figure \ref{Figure:topic_variation}, bottom row): there is no significant difference in AUC, and we only need $2$ participating groups to achieve a $\tau_{ap}$ correlation of 0.9 or above. Since NDCG@10 is far shallower than MAP@1000, it is reasonable to observe that we might need fewer groups to achieve $\tau_{ap} >= 0.9$ using NDCG@10. On the other hand, Max Drop for MAP and NDCG shows a noticeable difference in AUC, when we down-sample topics. 
We do not see any decreasing pattern in AUC for MAP or NDCG as we increase the number of topics.   

In these experiments, we see that for any given number of participating groups, increasing the number of topics does not improve test collection  reusability. For both $\tau_{ap}$ and Max Drop metrics, this observation holds.  However, we should acknowledge the limitations of our experiments. Firstly, since we experiment using only Robust'04 (given its 249 topics), further analysis on other test collections would be needed. Secondly, our experiments only vary the number of topics above 50, so results with more spartan topic sizes might vary. Thirdly, we assume a fixed pool depth of $100$. Fourthly, Robust'04 contains only automatic runs; manual runs often find other unique relevant documents. Fifthly,  Robust'04 dataset is relatively small compared to a modern collection such as ClueWeb'12 (Table \ref{table:group_runs_data}). Since larger collections tend to contain more relevant documents, increasing the number of topics for these larger document collections might be more valuable.

\begin{figure*}[h]
\centerline{\includegraphics[scale=0.40]{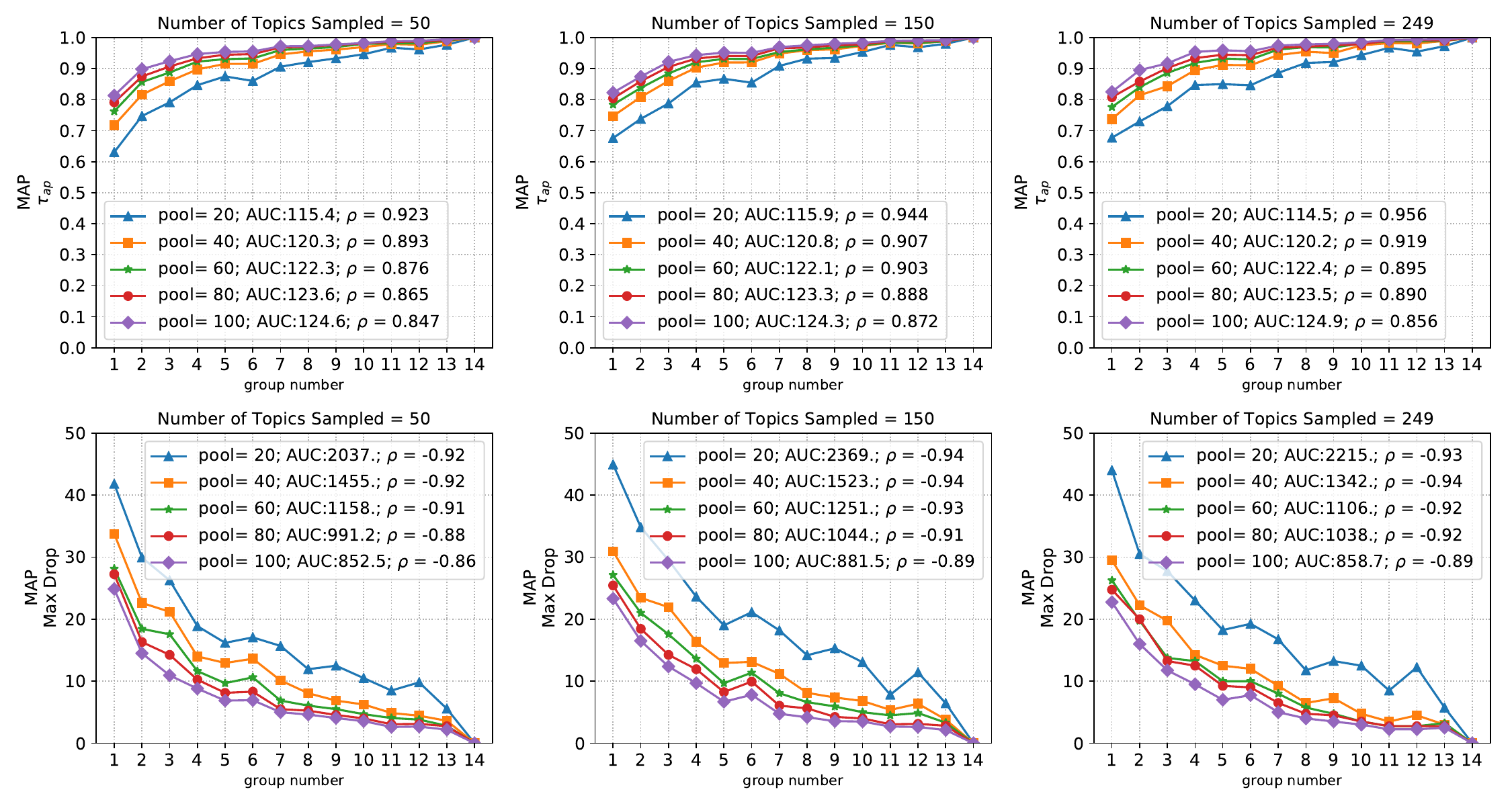}} 
\caption{$\tau_{ap}$ (first row), and Max Drop (second row) obtained by simulated test collections with a varying number of topics along with a varying pool depth on the Robust04 dataset. The x-axis represents the number of groups and the y-axis shows results when runs are ranked using MAP.}
\label{Figure:pool_variation}
\end{figure*}

\subsubsection{Impact of Pool Depth.} In the previous experiment, we observe how varying the number of topics interacts with varying the number of participating groups to build a reusable test collection while keeping the pool depth fixed. In this experiment, we also change the pool depth along with the number of topics and the number of groups. The experimental setup for this experiment is the same as discussed in the previous experiment except we vary the pool depth $p$ where $p$ takes values from the set $\{20, 40, 60, 80, 100\}$. 

The results for this experiment are reported in {\bf Figure \ref{Figure:pool_variation}}. The top row represents results for $\tau_{ap}$ whereas the bottom rows present results for Max Drop. The columns of Figure \ref{Figure:pool_variation} indicate the number of topics sampled for each of those pool depth variations. Here, we report the results where the runs are ranked using MAP. From the previous experiments, considering the limitations discussed above, we already know that increasing the number of topics does not improve the reusability of a test collection when the pool depth is 100. 
This still holds at each varying pool depth reported in Figure \ref{Figure:pool_variation}.  

By observing Figure \ref{Figure:pool_variation}, we find that for a fixed number of topics, increasing the pool depth improves the AUC in terms of $\tau_{ap}$ and lowers the value of AUC in terms of Max Drop, which indicates a better quality test collection. 
The greatest improvement in AUC happens when we increase the pool depth from 20 to 40 in all topic sets we investigate, suggesting that it has a high return on investment. Based on our results, if we have a large enough evaluation budget, using a pool depth of at least 40 is a reasonable choice. 

Another observation from Figure \ref{Figure:pool_variation} is  how the number of groups interacts with the pool depth. For example, in the plot for 50 topics, when we have a pool depth of only 20, we need at least 7 groups (half of the total number of groups in Robust'04 test collection), to achieve a $\tau_{ap}$ correlation of 0.9 or above. However, if the number of participants goes down to 3 groups (one-fifth of the total number of groups in Robust'04 test collection), we need a pool depth of $80$ to achieve the same $\tau_{ap}$ correlation. This observation holds for all of the other varying numbers of topics sampled in Figure \ref{Figure:pool_variation}. 

Based on the above discussion, we conclude that 
the number of participants is the most important factor for the quality of test collections. However, if we have few participating groups, rather than increasing the number of topics, we should increase 
pool depth in order to produce a reusable test collection. This conclusion is also subject to the same limitations discussed in the previous experiment, except for the pool depth. 

\begin{figure*}[!htb]
\centerline{\includegraphics[scale=0.50]{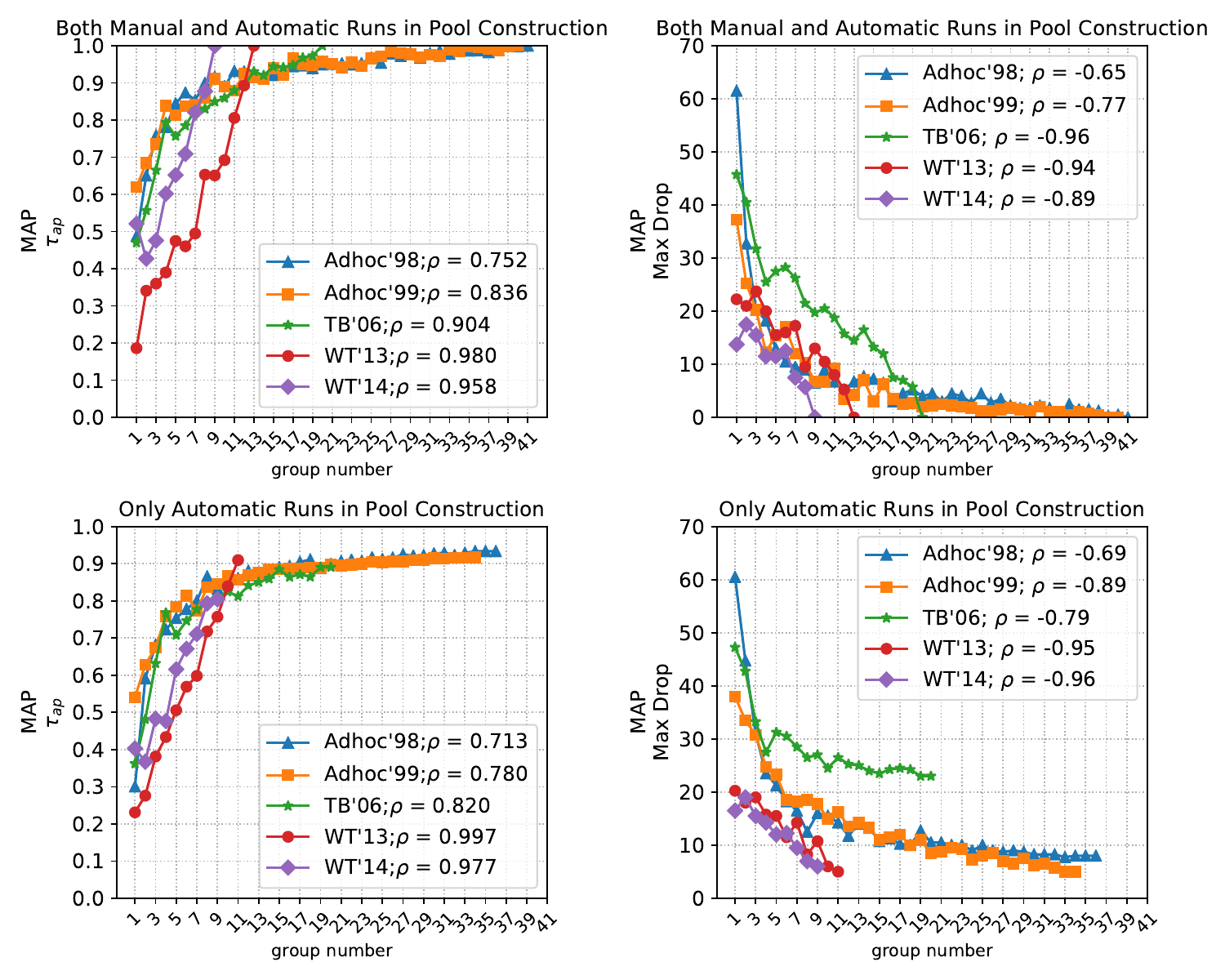}} 
\caption{ $\tau_{ap}$, and Max Drop obtained by simulated test collections with a varying number of groups with manual runs (top row) and without manual runs (bottom row) on the five TREC datasets. The x-axis represents the number of groups. Runs are ranked using MAP.}
\label{Figure:collection_variation}
\end{figure*}

\begin{figure*}[h]
\centerline{\includegraphics[scale=0.65]{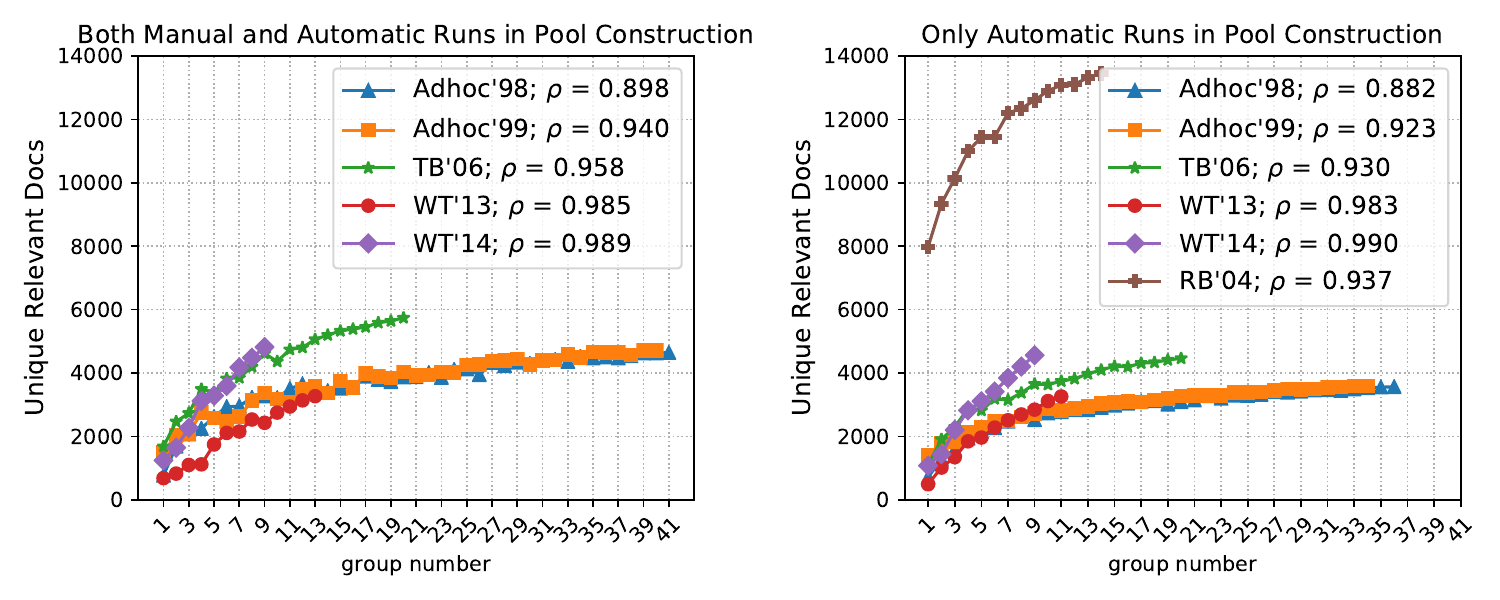}} 
\caption{ The number of unique relevant documents obtained by simulated test collections on six different test collections. Left plot does not include Robust'04 (RB'04) dataset as it contains only automatic runs. In the right plot, Robust'04 test collection  has a higher number of unique relevant documents because it has 249 topics whereas the other datasets have 50 topics only.}
\label{Figure:unique_document_all_collections}
\end{figure*}


\subsubsection{Impact of the Document Collection Size.} 
We conduct experiments on five different test collections, namely Adhoc'99, Adhoc'98, TB'06, WT'13, and WT'14, which 
contain both types of runs (i.e., manual and automatic runs).  We vary the number of groups and report respective $\tau_{ap}$, Max Drop scores (See {\bf Figure \ref{Figure:collection_variation}}), and the number of relevant documents (See {\bf Figure \ref{Figure:unique_document_all_collections}}), following Algorithm \ref {algorithm:ourmethod_I}. 
We do not report AUC 
since the AUC is not comparable across different collections.
The top row of Figure \ref{Figure:collection_variation} considers both manual and automatic runs in the construction of a test collection, while the bottom row only considers the automatic runs. We present results here using only MAP.

From Table \ref{table:group_runs_data}, we can see that the size of document collection for Adhoc'99 and Adhoc'98 test collection is very small ($\approx 0.5$ Million) compared to the size of document collection ($\approx 52$ Million) for WT'13 and WT'14 test collections. Although the number of unique relevant documents found depends on the number of topics, pool depth, and the number of participating groups, it is intuitive that for a fixed number of topics and pool depth, a larger document collection usually has a higher number of unique relevant documents. 
Therefore, the size of the document collection coupled with a varying number of participating groups will affect the reusability of a test collection. By observing Figure \ref{Figure:collection_variation} (top row), we can see that for Adhoc'98 and Adhoc'99 collections 
with both manual and automatic runs, we can achieve a $\tau_{ap}$ of 0.9 or above when there are 8 and 9 participating groups, respectively (i.e., approximately $20\%$ of the original number of participating groups for the respective test collections). However, for WT'13 and WT'14 datasets, we need 12 and 8 groups (Figure \ref{Figure:collection_variation}, top row), respectively, which is around on average $90\%$ of the original number of participating groups for those respective test collections.  

Although there is no acceptable range of values for Max Drop, if we assume that only $90\%$ of the original number of groups participate in WT'13 and WT'14 test collections, 
the Max Drop lies between 6 -- 10. Regarding Pearson correlation ($\rho$) scores computed between the number of groups and the corresponding $\tau_{ap}$ and Max Drop, we observe higher Pearson correlation values for WT'13 and WT'14 collections than with other collections. This is because 
$\tau_{ap}$ keeps increasing and Max Drop keeps decreasing with an increasing number of groups in WT'13 and WT'14 test collections. On the other hand, both Adhoc'99 and Adhoc'98 collections have low Pearson correlation scores because $\tau_{ap}$  increases and Max Drop decreases when the number of groups is increased from 2 to 10. However, after having 10 groups, $\tau_{ap}$ and Max Drop scores become almost stable for these two collections. 

In summary, we confirm that more participating groups are needed to have a reusable collection if the underlying document collection is very large. This likely stems from runs for these larger document collections returning more unique-relevant documents than runs in other test collections (Figure \ref{Figure:collection_variation}). Therefore, a run might be highly affected if it does not contribute to the pool. As another hypothesis, the groups participating in the recent shared-tasks are able to develop a more diverse set of IR runs than the groups participated in earlier shared-tasks due to the progress in the field of IR. 

However, it should be noted that the finding reported in this experiment has certain limitations. Firstly, all five test collections used in this analysis have only 50 topics. Secondly, for WT'13 and WT'14 collections, the employed pool depth is very shallow (Table \ref{table:group_runs_data}). A deeper pool depth and a higher number of topics might provide us a different conclusion than the one stated here.

\subsubsection{Impact of Manual Runs.}

In order to see the impact of manual runs on test collection reusability, we remove all manual runs from the simulated test collections and conduct the same experiments as  described above. \textbf{Figure \ref{Figure:collection_variation}}, bottom row presents the results for this particular setup. Note from {\bf Figure \ref{Figure:unique_document_all_collections}} that the number of unique-relevant documents is reduced noticeably when we do not use any manual run in AdHoc'98, AdHoc'99, and TB'06 test collections, confirming the importance of manual runs providing more unique-relevant documents than automatic runs. We also observe that WT'13 and WT'14 test collections are less affected than other collections in terms of the number of unique relevant documents because there are fewer manual runs in these test collections (See Table \ref{table:group_runs_data}).

Comparing results between the top and bottom rows of Figure \ref{Figure:collection_variation}, we find that not having any manual run greatly increases the required number of participating groups to achieve a $\tau_{ap}$ correlation of 0.9 or above. For example, for Adhoc'98, and Adhoc'99 collections, we need at least $50\%$ of the original number of participants (Table \ref{Figure:collection_variation}, bottom row) to achieve a $\tau_{ap}$ correlation of 0.9 or above which is only $20\%$ (Table \ref{Figure:collection_variation}, top row) when we do include manual runs. For TB'06 and WT'13 collections, we actually need $100\%$ of the original number of participants. 
Our observations are also similar for Max Drop scores.  This suggests that the unique-relevant documents detected by manual runs could not have been detected by any of the automatic runs, thereby affecting the ranking of manual runs during the evaluation. 

Ultimately, if we have very few participating groups, it is especially important to have manual runs to develop a reusable test collection. Note that this finding  is based on 
an experiment with  50 topics and a shallow pool depth  for the WT'13 and WT'14 test collections. \\ 

\begin{table*}[h]
\vspace{-1em}
\centering
\caption{Performance of the multiple linear regression model using Leave-one-test-collection-out setup on five different test collections.}
\begin{tabular}{|c|c|c|c|}
\hline
{\bf Training} & {\bf Testing} & {\bf Intersection of} &{\bf MSE}\\
{\bf Set} & {\bf Set} & {\bf Document Collection} &{\bf $\tau_{ap}$} \\
&  & {\bf between Train \& Test Set} & \\
\hline
Adhoc'99 , TB'06 , WT'13 , WT'14 & Adhoc'98 & Yes & 0.005\\
Adhoc'98 , TB'06 , WT'13 , WT'14 & Adhoc'99 & Yes & 0.003\\
Adhoc'98 , Adhoc'99 , WT'13 , WT'14 & TB'06 & No & 0.171 \\
Adhoc'98 , Adhoc'99 , TB'06 , WT'14 & WT'13 & Yes & 0.060\\
Adhoc'98 , Adhoc'99 , TB'06 , WT'13 & WT'14 & Yes & 0.055\\
\hline
\end{tabular}
\label{Table:regression_model}
\end{table*}

\vspace{-2em}
\section{Predicting the Qualities of Test Collections}
\label{sec_exp_2}

We investigate whether it is possible to forecast the quality of a test collection  before gathering the ranked lists of participants. Our rationale is that the shared-task organizers can act accordingly based on the predicted quality of a test collection even before spending budget on collecting relevance judgments. In this study, we focus on predicting $\tau_{AP}$ as a measure of reusability. 

To generate data for our model, we use the same simulated test collections constructed from Algorithm \ref{algorithm:ourmethod_I} and employ MAP to compute $\tau_{ap}$. We utilize the following features for the prediction model: i) the number of participating groups ($G$), ii) the number of topics ($T$), iii) the pool depth ($P$) and iv) the size of the document collection ($C$). Then we fit a Multiple Linear Regression model: $\hat{y} = W_0 + W_1*G + W_2*T + W_3*P + W_3*C$ on the training data to predict $\tau_{ap}$. Here W's are the learned weights for the features of the model. As a performance measure of our prediction model, we report  Mean Squared Error (MSE) $=  \sum_{i=1}^n\frac{(\hat{y_i} - y_i)*(\hat{y_i} - y_i)}{n}$,
where $y_i$, and $\hat{y_i}$ are the predicted and true target value of $\tau_{ap}$, respectively, and $n$ is the total number of data points. A lower value of MSE indicates a better model. 

To understand the generalization performance of the prediction model, we employ ``\textit{Leave-one-test-collection-out}" (LOTO) strategy. In this LOTO setup, in turn, we hold out one test collection and utilize the remaining test collections from the set of test collections to train the prediction model, and then we test the predictive performance of the trained model on the held-out test collection. For example, in the first row of \textbf{Table \ref{Table:regression_model}}, we utilize Adhoc'99, TB'06, WT'13, and WT'14 test collections as training set and Adhoc'98 test collection as testing set for our model. Since Adhoc'98 in the test set shares the same document collection (Table \ref{table:group_runs_data}) with Adhoc'99 from the train set, the 3rd column of 1st row indicates ``Yes". In contrast, TB'06 in the testing set (3rd row) does not share the document collection with any of the test collections utilized in the training set and thus the 3rd column of the 3rd row indicates ``No".
MSE for predicting $\tau_{ap}$ is reported in 4th column of Table \ref{Table:regression_model}.

From Table \ref{Table:regression_model}, we observe that the model can predict $\tau_{ap}$ with a very high accuracy (MSE $\le 0.06$) for all five test collections except for TB'06. This may be because a) runs in Adhoc'98 and Adhoc'99 might be similar due to closeness of two shared-tasks in terms of time, and b) both have the same document collection (Column 3 of Table \ref{Table:regression_model}). We can apply the same reasoning for WT'13 and WT'14 test collections. 


In summary, our prediction performance improves when we use test collections from the same document collection for both training and testing sets of the model. In practice, our prediction model can be especially useful after its first year as TREC usually continues a track for more than one year. Thus, the results in the first year can be used to forecast the quality of the test collections in the following years for the same track.  Test collection construction parameters (i.e., topic set size and pool depth) can be set based on the predictions.


\vspace{-0.5em}
\section{Conclusion and Future Work}\label{sec_conc}
\vspace{-0.5em}

In this work, we investigate how varying the number of participating groups coupled with other factors in a shared-task affects the reusability of the resulting test collection.  
Our main findings based on experiments conducted on six TREC test collections are as follows. Firstly, when we have very few participating groups in a shared-task, increasing the pool depth provides a more reusable test collection than increasing the number of topics. Secondly, the size of the document collection and the types of runs play a crucial role when the number of participants is very small. Thirdly, we show that the reusability of a test collection can be predicted with high accuracy when the same document collection is used for successive years in an evaluation campaign, as is commonly done.


There are many possible future directions for this work. For example, our experimental analysis is conducted on pooling-based test collections. However, we plan to extend our work on analyzing test collections constructed using other techniques. For example,  the TREC 2017 Common Core track \citep{allan2017trec} test collection is constructed using a bandit-based technique \citep{losada_feeling_2016}. In addition, we also plan to address the mentioned limitations of  our analysis by using more test collections. 
Future work might also explore the prediction of test collection quality when the ranked lists of documents are submitted but relevance judgments are not collected yet. This scenario will enable using different features such as the number of unique documents retrieved by groups, rank correlation of ranked lists, and others with more sophisticated machine learning models.

\textbf{Acknowledgments}. We thank the reviewers for their valuable feedback. This research was supported in part by Wipro, the Micron Foundation, and by Good Systems\footnote{\url{http://goodsystems.utexas.edu/}}, a UT Austin Grand Challenge to develop responsible AI technologies. The statements made herein are solely the opinions of the authors and do not reflect the views of the sponsoring agencies.

\bibliographystyle{splncsnat}
\bibliography{bibliography}

\end{document}